\documentclass[5p]{elsarticle}

\usepackage{lineno,hyperref}
\modulolinenumbers[5]

\usepackage{amsmath}
\usepackage{amssymb}

\usepackage{graphicx}
\usepackage{subfig}
\graphicspath{{figures/}}

\journal{Acta Astronautica}

\begin{document}

\begin{frontmatter}

\title{The antenna phase center motion effect in space-based experiments for fundamental physics and astronomy}



\author[sai]{A.~I.~Filetkin}

\author[sai,asc,bmstu]{D.~A.~Litvinov\corref{mycorrespondingauthor}}
\cortext[mycorrespondingauthor]{Corresponding author at: Sternberg Astronomical Institute, Lomonosov Moscow State University,
Universitetsky pr.~13, 119991 Moscow, Russia.}
\ead{litvirq@yandex.ru}

\author[sai]{V.~N.~Rudenko}

\author[kiam]{M.~V.~Zakhvatkin}

\address[sai]{Sternberg Astronomical Institute, Lomonosov Moscow State University,
Universitetsky pr.~13, 119991 Moscow, Russia}

\address[asc]{Astro Space Center, Lebedev Physical Institute, Profsoyuznaya 84/32, 117997 Moscow, Russia}

\address[bmstu]{Bauman Moscow State Technical University, 2-ya Baumanskaya 5, 105005 Moscow, Russia}

\address[kiam]{Keldysh Institute for Applied Mathematics, Russian Academy of Sciences, Miusskaya sq. 4, 125047 Moscow, Russia}

\begin{abstract}

We consider the effect of phase center motion of mechanically steerable high-gain parabolic antennas for ground-based and spacecraft-mounted antennas.  For spacecrafts on highly elliptic Earth orbits the magnitude of the effect is as large as several mm/s in terms of the required velocity correction, both for ground-based and spacecraft-mounted antennas. We illustrate this with real data from the RadioAstron spacecraft and also provide results of our simulations for the concept of a possible follow-up space very long baseline radio astronomy mission. We also consider a specific configuration of satellite communication links, with simultaneously operating one-way down link and two-way loop link, pioneered by the Gravity Probe A experiment. We find that this configuration provides for complete compensation of the phase center motion effect due to the onboard antenna and significant compensation of that due to the ground antenna. This result is important for future space-based fundamental physics experiments, primarily those concerned with studies of gravity.
\end{abstract}


\end{frontmatter}


\section{Introduction}

The effect of phase center motion of steerable high-gain antennas has long been recognized as an important part of time-delay models of very long baseline radio interferometry (VLBI) \cite{wade-1970-apj} and the source of a significant correction to computed values of observables in spacecraft orbit determination \cite{moyer-1971-techreport,moyer-2005-book}. In both cases one is usually concerned with ground-based high-gain mechanically steerable dish antennas, several tens of meters in diameter, and the correction is due to a non-zero offset between the rotation axes of the antenna mount. Spacecraft-mounted antennas also exhibit this effect but, up until recently, steerable high-gain antennas have been mostly used by deep space probes and other planet orbiters, in which case the magnitude of the effect is generally considered negligible due to large distance to the Earth \cite{lemoine-2001-jgr}.

The success of the first space-VLBI missions, VSOP \cite{hirabayashi-1998-sci} and RadioAstron \cite{kardashev-2013-ar}, utilizing Earth-orbiting satellites with high-gain mechanically steerable dish antennas, makes it necessary to reconsider the importance of the spacecraft part of the antenna effect, which we attempt with this paper. In space-VLBI the necessity for a high-gain steerable onboard antenna comes from the requirement of high data transmission rate (up to several Gbits/sec) and high eccentricity of the orbit ($e > 0.5 $). Frequency measurements of the data downlink signal are input into the orbit determination and thus should be corrected for the antenna motion effect. We present the results of real data processing from the RadioAstron spacecraft and demonstrate the importance of taking into account the phase center motion both for ground and on-board antenna. We find that the magnitude of the effect is as large as 6 mm/s in terms of the velocity correction and thus must be taken into account not only for precise but even ordinary orbit determination. These demonstrate that these results are relevant also for future space-VLBI missions, such as \cite{kardashev-2014-ufn} and \cite{hong-2014-aa}.

We also consider a specific configuration of spacecraft radio communication links, namely, that of simultaneously operating one-way downlink and two-way loop link. Such configuration was first used by the Gravity Probe A mission \cite{vessot-levine-1980-prl} for first-order Doppler and tropospheric frequency shift compensation and is now an integral part of mission concepts for space-VLBI and space-based fundamental physics experiments \cite{altschul-2015-asr, jetzer-2017-ijmpd}. We show that in the output of this “compensation scheme” the antenna phase center effect of the one-way downlink and that of the loop link cancel each other, both for the ground- and onboard antenna. The cancellation depends on the distance to the spacecraft, angular velocity of the spacecraft
relative to the ground station and the rate of change of the angular velocity. This observation provides for effectively using high-gain mechanically steerable dish antennas in high-precision and multidisciplinary space-based physics missions.

The antenna phase center motion of non-steerable antennas has recently gained much attention in the problem of precise orbit determination of satellites of global navigation satellite systems (GNSS) \cite{springer-2017-handbook}. In this case the phase center offset and its variation is routinely taken into account not only for ground-based receiving antennas but also for transmitting antennas of GNSS satellites and GNSS-receivers of other satellites. Although this is a similar effect, we do not consider it in this paper.

The outline of the paper is as follows. In Section 2 we summarize the equations for the antenna effect for ground-based steerable dish antennas and present the equations for taking into account the effect for onboard steerable dish antennas. In Section 3 we present the results of data processing from the RadioAstron satellite and demonstrate the importance of taking into account the ground- and onboard antenna phase center motion for regular (not precise) orbit determination. In Section 4 we consider the Chinese mission concept of a space-VLBI mission with two spacecrafts in highly elliptic Earth orbits and show the importance of taking into account the antenna effect for this case. In Section 5 we consider the Gravity Probe A configuration of radio communication links and show that in this case the antenna effects cancel. We conclude with a discussion in Section 6.

\section{Basic Equations}

\begin{figure}[h!]              
        \centering
        \subfloat[Alt-azimuth mount]{
                \label{fig:fig1:a}
                \includegraphics[scale=0.33]{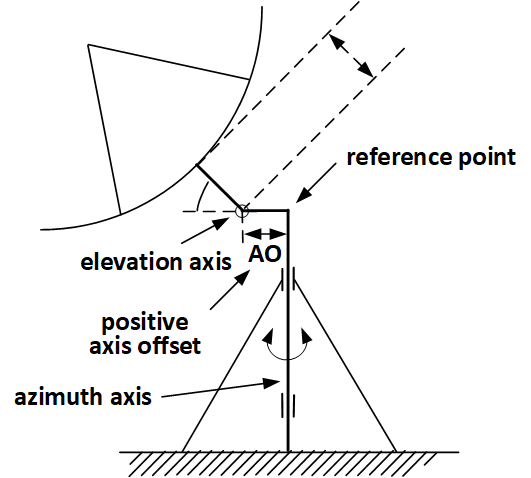}}
        \subfloat[Polar mount]{
                \label{fig:fig1:b}
                \includegraphics[scale=0.33]{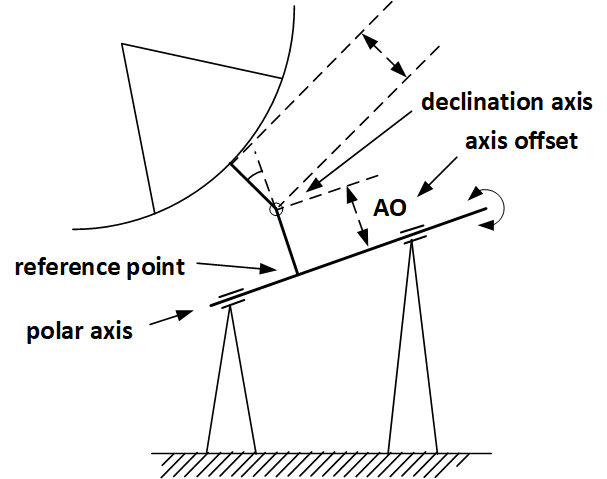}}
        
        \subfloat[X-Y mount]{
                \label{fig:fig1:c}
                \includegraphics[scale=0.33]{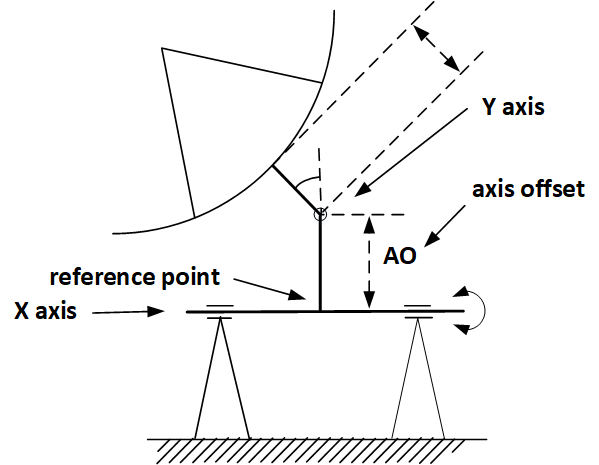}}
        \caption{Ground antenna mounts.}
        \label{fig:fig1}
\end{figure}

The antenna phase center motion effect for mechanically steerable dish antennas is due to motion of the antenna phase center relative to a fixed reference point. In order to define this reference point it is necessary to consider the existing types of antenna mounts. Four types of steerable ground antenna mounts are in common use (Fig.~\ref{fig:fig1}). The most common is the alt-azimuth mount with a fixed primary azimuth axis and the secondary moving axis being the elevation axis (Fig.~\ref{fig:fig1:a}). Polar, or equatorial, mount antennas have their polar axes oriented in parallel to the Earth’s rotation axis (Fig.~\ref{fig:fig1:b}), which provides for easy tracking of celestial objects with just a single rotation of the polar axis. For X-Y mount antennas the fixed axis lies in the local horizontal plane, and is orients in the North-South or East-West direction (Fig.~\ref{fig:fig1:c}). For many antennas the primary and secondary axis do not intersect, with the offset varying from a few millimeters to a few meters. The most prominent example is the Green Bank 43-meter polar mount radio telescope, for which the offset is 14.9 m.

\begin{figure}[h]               
        \centering
        \includegraphics[scale=0.5]{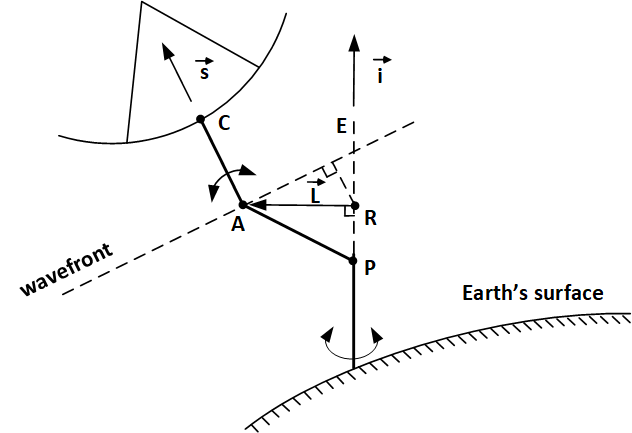}
        \caption{Generalized schematic representation of the geometry of a steerable antenna.}
        \label{fig:fig2}
\end{figure}

For antennas with intersecting rotation axes of the mount the antenna reference point is defined as the intersection of its rotation axes. For antennas with non-intersecting axes the reference point is defined as the orthogonal projection of the secondary axis onto the primary one \cite{moyer-2005-book}.
In order to obtain equations for the antenna phase center offset we follow the treatments of \cite{wade-1970-apj}, \cite{sovers-1998-rmp} and \cite{sovers-1987-techreport}. Consider an incoming plane wave, propagating in the direction of the dish along its symmetry axis (Fig. \ref{fig:fig2}) (taking into account wavefront curvature gives a negligible correction \cite{sovers-1998-rmp}) . This wave will reach the antenna phase center (point C) earlier than the Earth-fixed antenna reference point R by 
$$\tau = l/c.$$
Now we define:
1)      a unit vector $\mathbf{i}$ along the primary axis (for alt-az and polar mounts in the direction away from the Earth, for X-Y mounts in the direction from North to South or from East to West, depending on the orientation of the primary axis);
2)      unit vector $\mathbf{s}f$ in the direction of the source;
3)      axis separation distance $L$.
Then one can easily see (\cite{sovers-1998-rmp}) that:
$$ l = \pm L \sqrt{1 - (\mathbf{s} \cdot \mathbf{i})^2},$$
where the plus or minus sign is chosen to give $\mathbf{L}$ the direction from R to A. The plus sign is used when the point A comes closer to the source than R.
The particular expressions for $l$ for different antenna mounts are given in Table \ref{table:table1}. The extra delay is obviously the same in sign and magnitude both for reception and transmission.

The corresponding correction to the frequency of the received or transmitted signal is, obviously,
\begin{equation}        \label{eq:eq1}
        \frac{\Delta f}{f} = \pm \frac{L \sin\theta(t_{i})\cdot\dot\theta(t_{i})}{c},
\end{equation}
where $\theta$ is the angle between vector $\mathbf{L}$ and $\mathbf{s}$, $\dot\theta$ is angle derivative at the moment of signal transmit or receive  $t_{i}$, $c$ is the speed of light,
which is also given in Table~\ref{table:table1} for each mount. 

\begin{table*}[h!]              
        \centering
        \begin{tabular}{|c|c|c|c|} \hline
                Antenna mount& Secondary angle & Delay & Frequency correction\\ \hline
                alt-az & elevation $ \gamma $ & $ l = \pm L \cos{\gamma} $ & $\dfrac{\Delta f}{f} = \dfrac{L \sin\gamma(t_{i})\cdot\dot\gamma(t_{i})}{c}$ \\
                polar & declination $ \delta $ & $ l = \pm L \cos{\delta} $ & $\dfrac{\Delta f}{f} = \dfrac{L \sin\delta(t_{i})\cdot\dot\delta(t_{i})}{c}$ \\
                X-Y & auxiliary angle $ Y $ & $ l = \pm L \cos{Y} $ & $\dfrac{\Delta f}{f} = \dfrac{L \sin Y(t_{i})\cdot\dot Y(t_{i})}{c}$ \\ \hline
        \end{tabular}
        \caption{Particular expressions for different antenna mounts.}
        \label{table:table1}
\end{table*}

Let us now consider the case of a spacecraft-mounted steerable dish antenna.
\begin{figure}[h!]              
        \centering
        \includegraphics[scale=0.4]{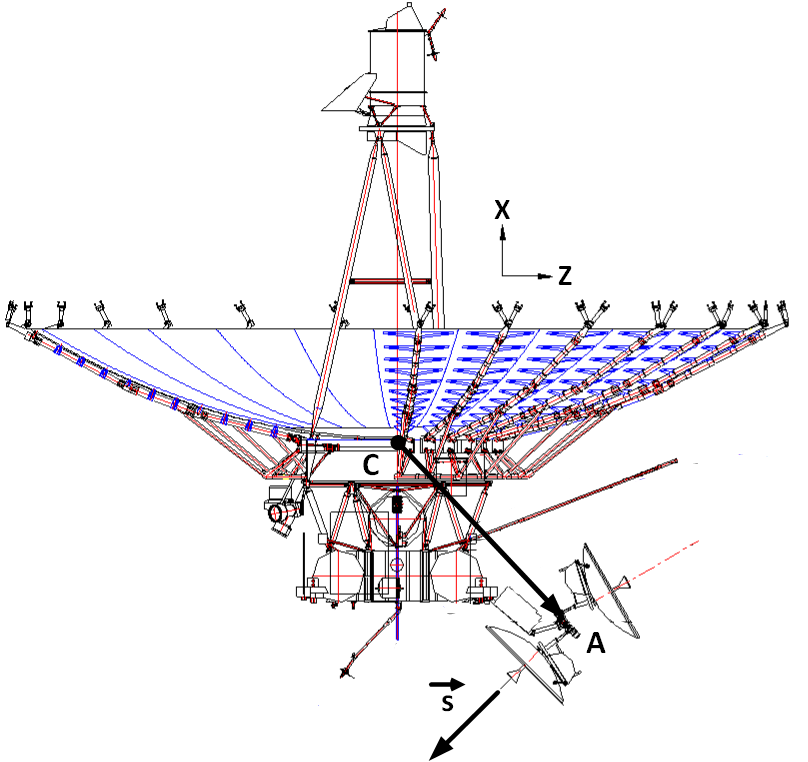}
        \caption{The scheme of the RadioAstron satellite \cite{fedorchuk-2014-cosres}. The displacement of the phase center relative to the center of mass of the satellite is defined as $\mathbf{b} = CA$.}
        \label{fig:fig3}
\end{figure}
The orbit of a spacecraft is described as the trajectory of the motion of a material point corresponding to the center of mass of the satellite. High-gain antenna phase center has a fixed position in spacecraft frame and denoted by $\mathbf{b}$. Unit vector $\mathbf{s}$ in the direction of the source also defined in spacecraft frame, taking into account its orientation in inertial reference system.
The delay of the received or transmitted signal equals $$ l = (\mathbf{b} \cdot \mathbf{s}). $$
The plus sign for delay is used because point A is always closer to the source than point C from spacecraft construction.
The corresponding correction to the frequency during tracking the source is

\begin{equation}        \label{eq:eq2}
\frac{\Delta f}{f} = -\frac{1}{c}\left(\mathbf{b} \cdot \dot{\mathbf{s}}(t_{i})\right)  ,
\end{equation}
where $\mathbf{b} = (x_{b}, y_{b}, z_{b})$ and $\dot{\mathbf{s}} = (\dot{x}_{s}(t_{i}), \dot{y}_{s}(t_{i}), \dot{z}_{s}(t_{i}))$.

\section{Application to RadioAstron}
\nopagebreak

The RadioAstron spacecraft, officially called Spektr-R, is an Earth-orbiting 10-meter radio telescope used primarily for VLBI observations of distant radio sources \cite{kardashev-2013-ar}. The satellite is on a highly eccentric orbit around the Earth, evolving
due to the gravitational influence of the Moon, as well as other factors, within a broad range of the orbital parameter space (perigee
altitude 1,000--80,000 km, apogee altitude 270,000--370,000~km). 

The satellite is equipped with a steerable 1.5~m high-gain parabolic antenna, used both for downlink and uplink communications with ground tracking stations (there are also telemetry and control low-gain antennas which we do not consider here). The rotation axes of the spacecraft antenna are offset from each other by several centimeters (Fig.~\ref{fig:fig3}), which we ignore here. The intersection of the rotation axes (point A) is offset from the center of mass (point C) by vector:
$$
\mathbf{b} = \left[ -2.298742, \: 0, \: 2.545722 \right]   ,
$$
with the X-coordinate depending on the amount of fuel in the tank.
The mission is served by two tracking stations \cite{kardashev-2013-ar,
ford-2014-procs}, with their parameters given in Table~\ref{table:table2}. 

\begin{table*}[h!]              
        \centering
        \begin{tabular}{|c|c|c|c|c|} \hline
                Tracking station & Location & Diameter & Mount & Axis offset \\ \hline
                Pushchino & Moscow region, Russia & 22 m & alt-az & 0 m \\
                Green Bank & West Virginia, USA & 43 m & polar & 14.94 m \\ \hline
        \end{tabular}
        \caption{Tracking stations.}
        \label{table:table2}
\end{table*}

\begin{figure}[h!]              
        \centering
        \includegraphics[scale=0.33]{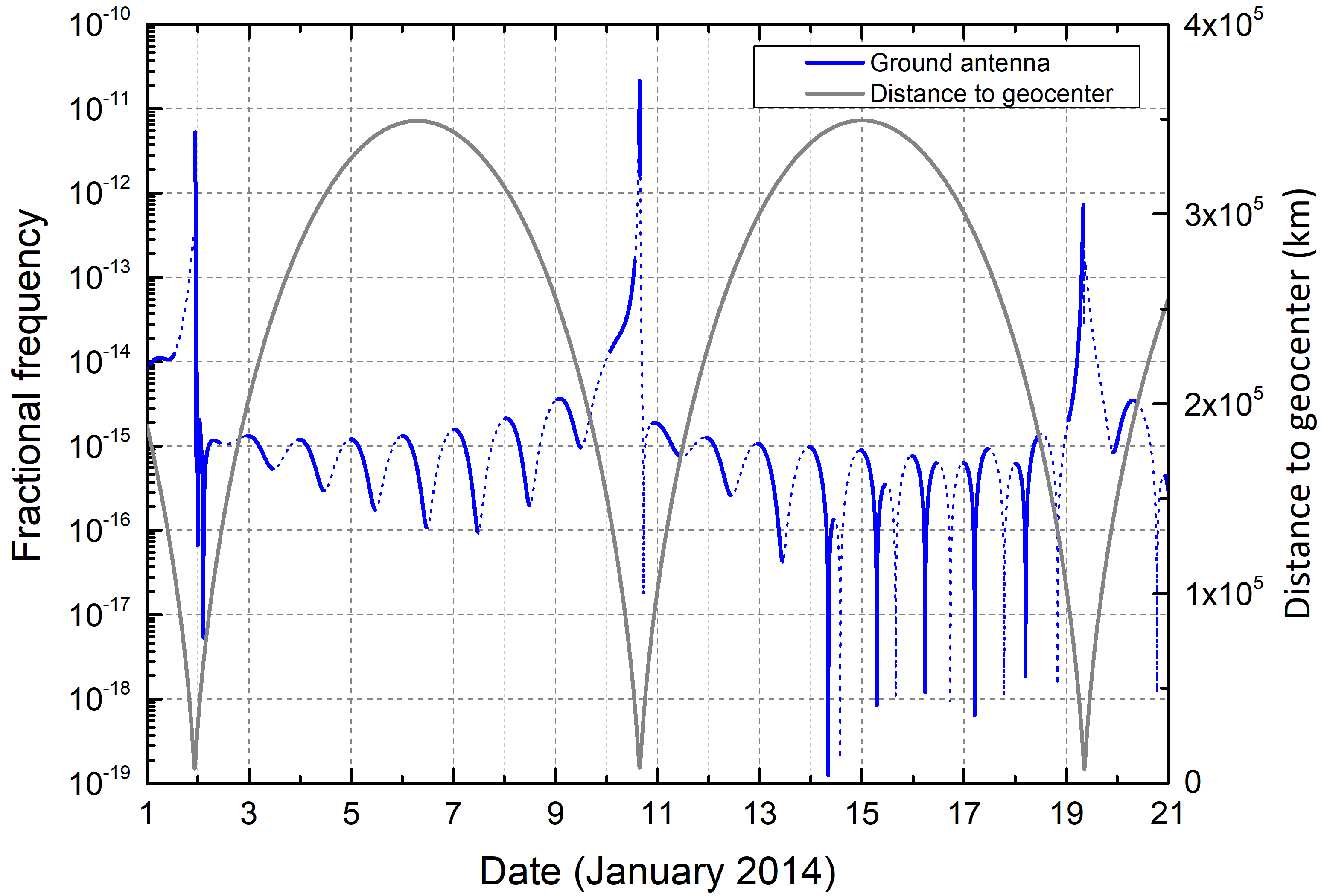}
\caption{Variation of fractional frequency shift of downlink signal along the orbit due to phase center motion of the Green Bank tracking station antenna to the January, 2014. Blue dots indicate the time intervals when the satellite under the horizon for the station.}
\label{fig:fig4}
\end{figure}

One would expect the large axis offset of the Green Bank antenna, coupled with high angular velocity of the spacecraft near perigee passage, to produce significant effects on measured Doppler frequency. This is indeed the case, as seen in Fig.~\ref{fig:fig4}, which depicts the absolute value of the antenna correction to the downlink Doppler frequency for the period of January 2014. The magnitude of the effect near perigee reaches 6 mm/s, which is 2 orders
of magnitude larger than  0.02--0.1 mm/s RMS of a typical 60s-averaged velocity residuals \cite{zaslavsky-2016-la}. 

\begin{figure}[h!]              
        \centering
        \includegraphics[scale=0.33]{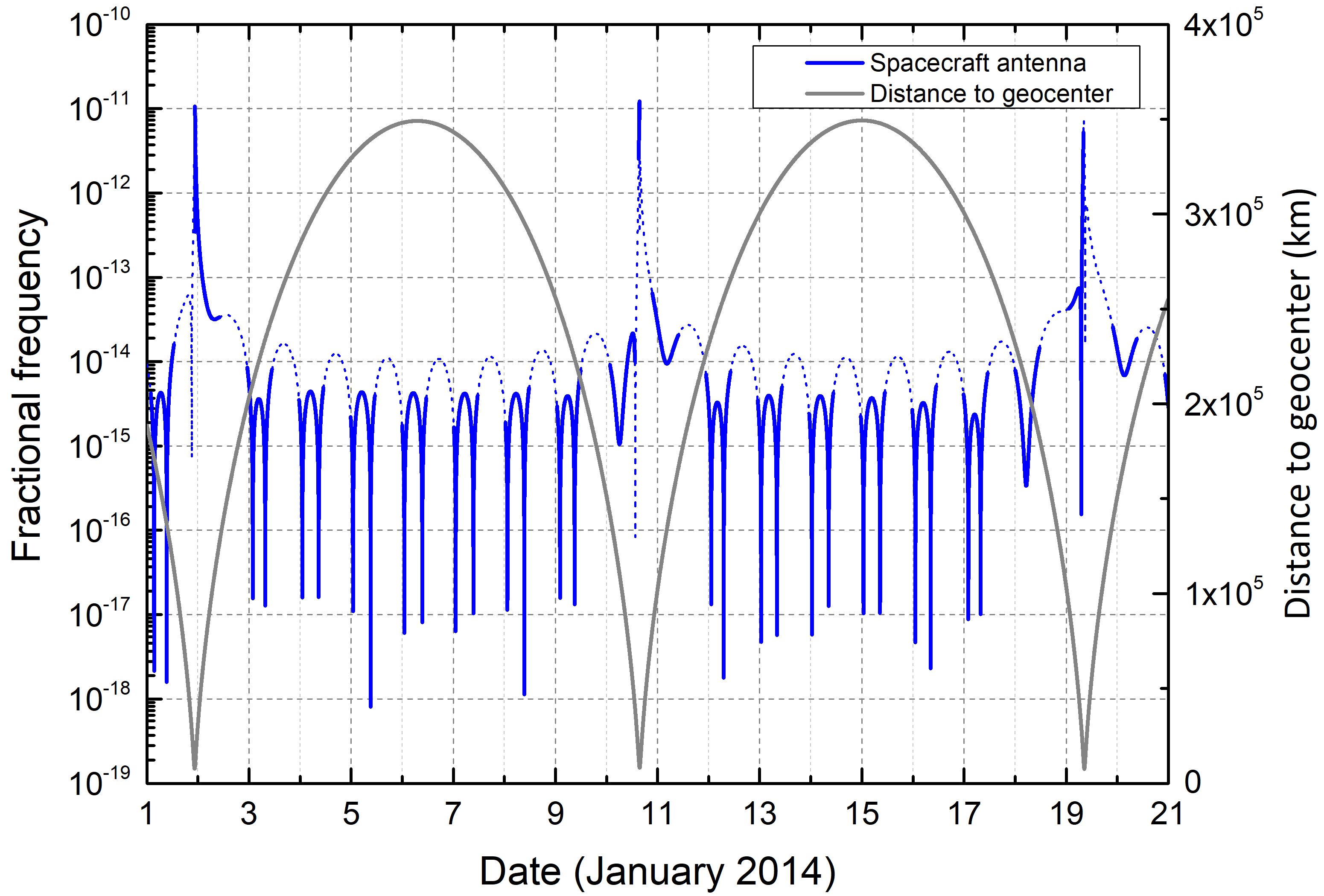}
\caption{Variation of fractional frequency shift of downlink signal along the orbit due to phase center motion of RadioAstron's on-board antenna in
the epoch of January 2014. Blue dots indicate the time intervals when the satellite was not visible by the tracking station.}
\label{fig:fig5}
\end{figure}

Consider now the effect of the 3.43-meter axis offset of the onboard antenna from the spacecraft center of mass. Fig.~\ref{fig:fig5} shows the magnitude of the required Doppler correction for the low-perigee period of January 2014. Clearly, the magnitude of the onboard antenna effect is almost as large as that for the ground antenna, reaching 1.1 mm/s on January 19, 2014.

\begin{figure}[htp]             
        \centering
        \subfloat[]{
                \label{fig:fig6:a}
                \includegraphics[clip, scale=0.33]{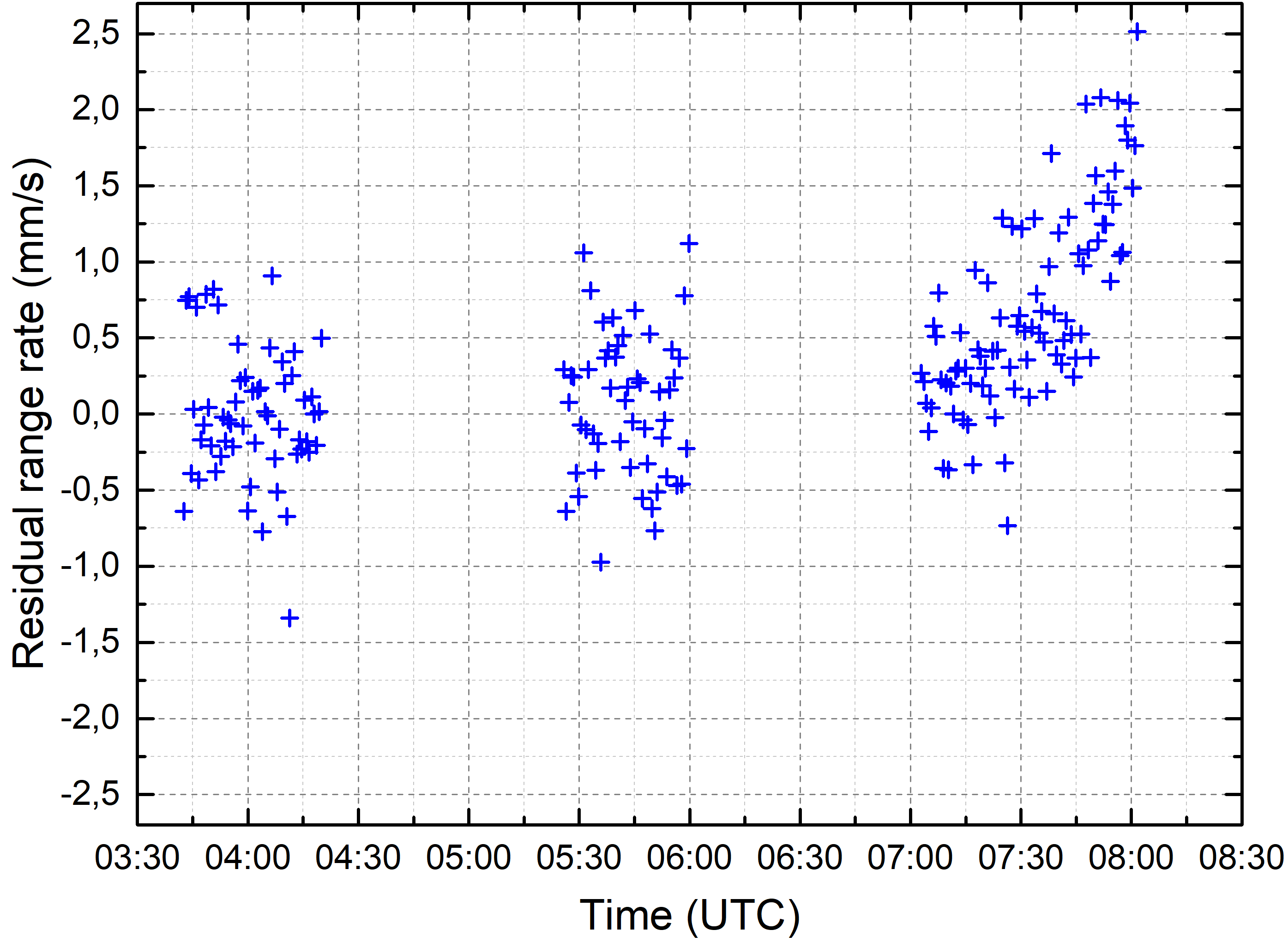}}
        
        \subfloat[]{
                \label{fig:fig6:b}
                \includegraphics[clip, scale=0.33]{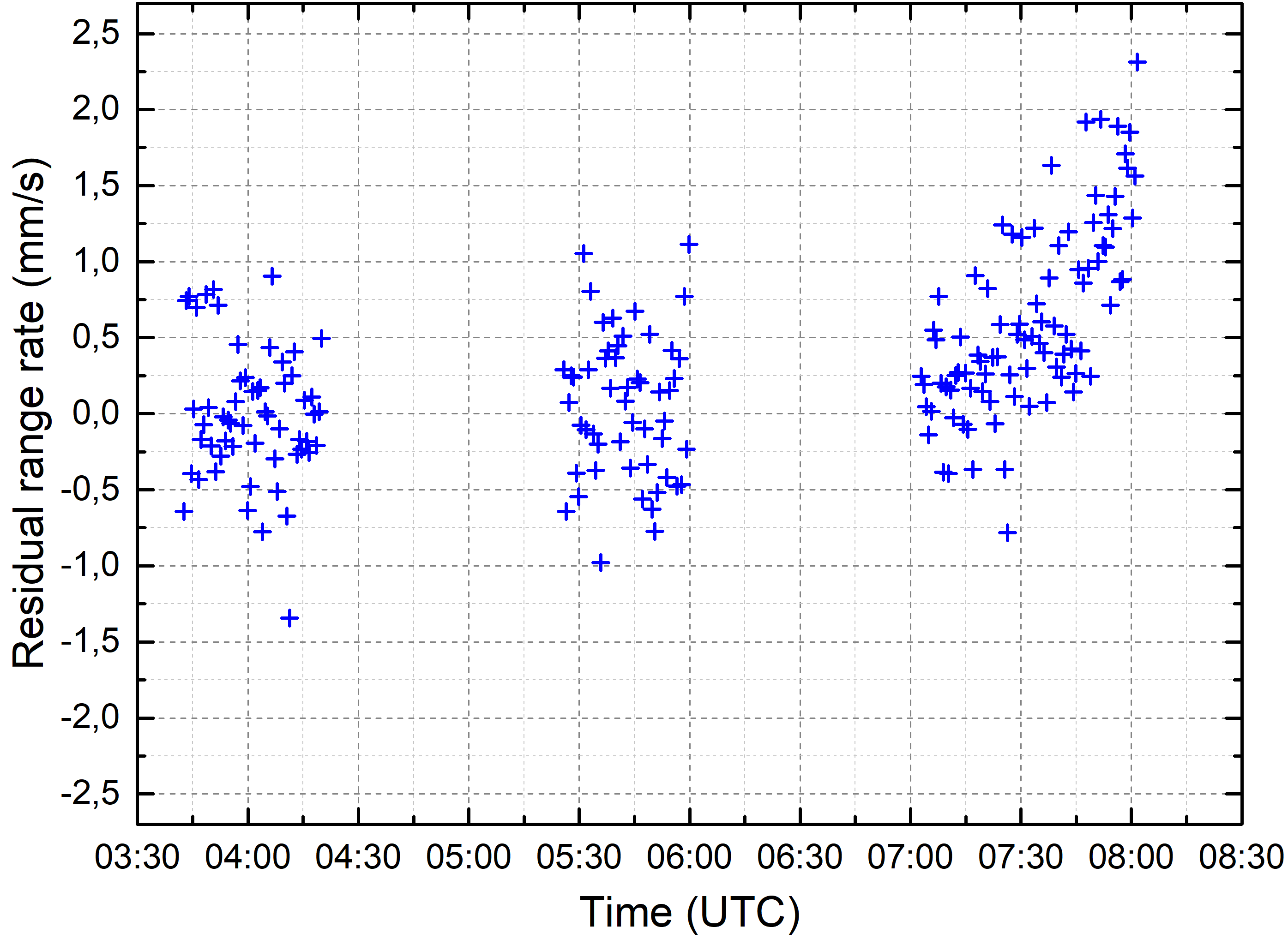}}
        
        \subfloat[]{
                \label{fig:fig6:c}
                \includegraphics[clip, scale=0.33]{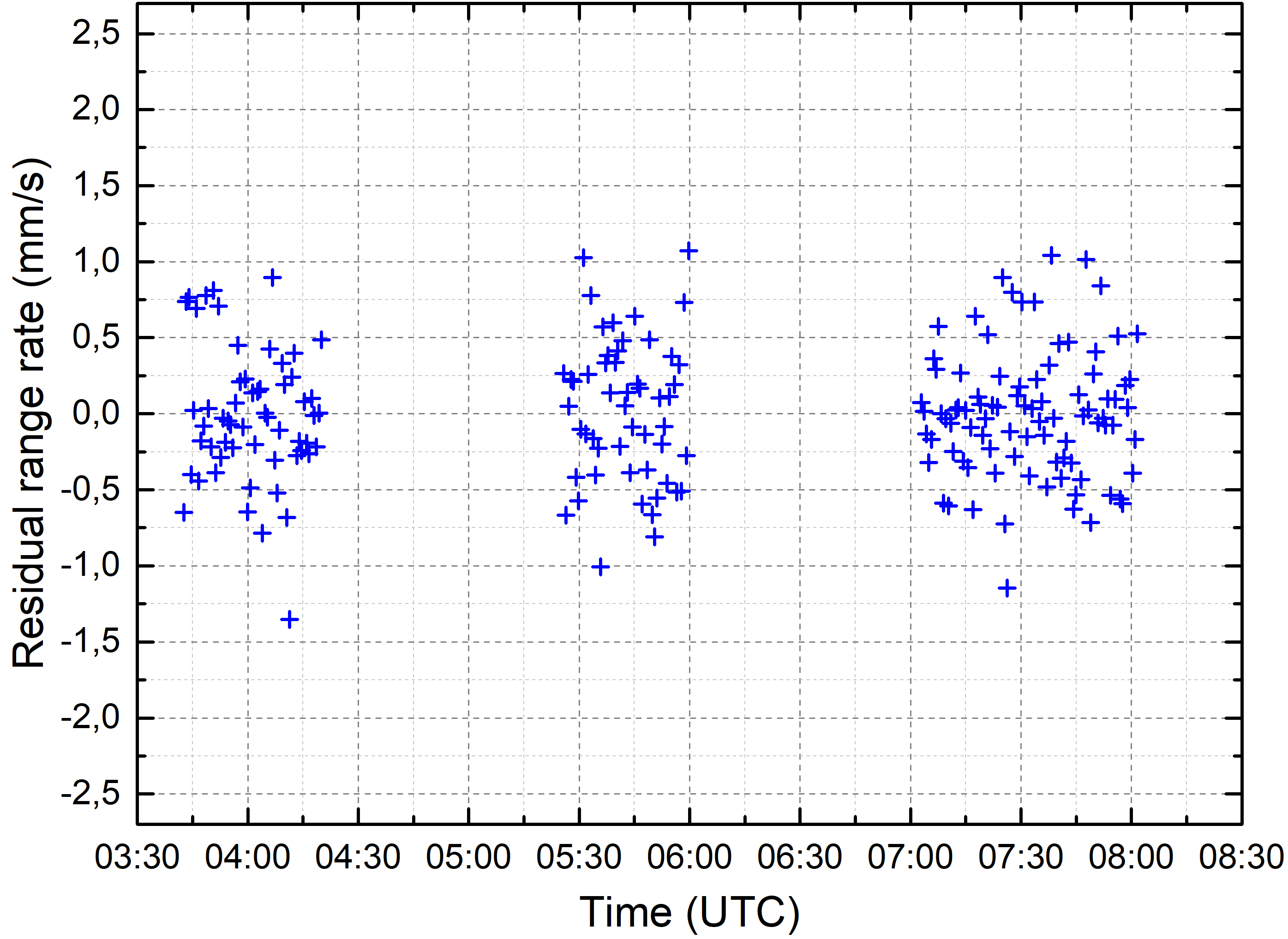}}
        \caption{One-way Doppler residuals during GB 2014-Jan-19. (a) - excluding onboard and ground antenna centers motions, (b) - excluding onboard antenna center motion, (c) - with considering both effects}
        \label{fig:fig6}
\end{figure}

Let us now look on how much important it is to take these effects into account in the problem of orbit determination.  The RadioAstron high-gain antenna transmits signals at 8.4 and 15 GHz routinely measured at tracking stations. These frequencies make an important part of the input to orbit determination algorithms of the spacecraft \cite{zaslavsky-2016-la}. Frequency measurements are recorded  with a time step of 0.04 s during each science observation with the space
radio telescope. Now we pick a particular observation during the low perigee epoch, 19/01/2014 03:50--08:00 UTC (experiment code raks04c) and plot the frequency residuals with the antenna effects not taken into account (Fig.~\ref{fig:fig6:a}), only the ground antenna effect taken into account (Fig.\ref{fig:fig6:b}) and both ground and onboard antenna phase center motion effects taken into account (Fig.~\ref{fig:fig6:c}). Clearly, taking into account the antenna phase center motion is very important for orbit determination.

\section{Application to future space-VLBI missions}
\nopagebreak
 
\begin{figure}[h!]              
        \centering
        \includegraphics[scale=0.33]{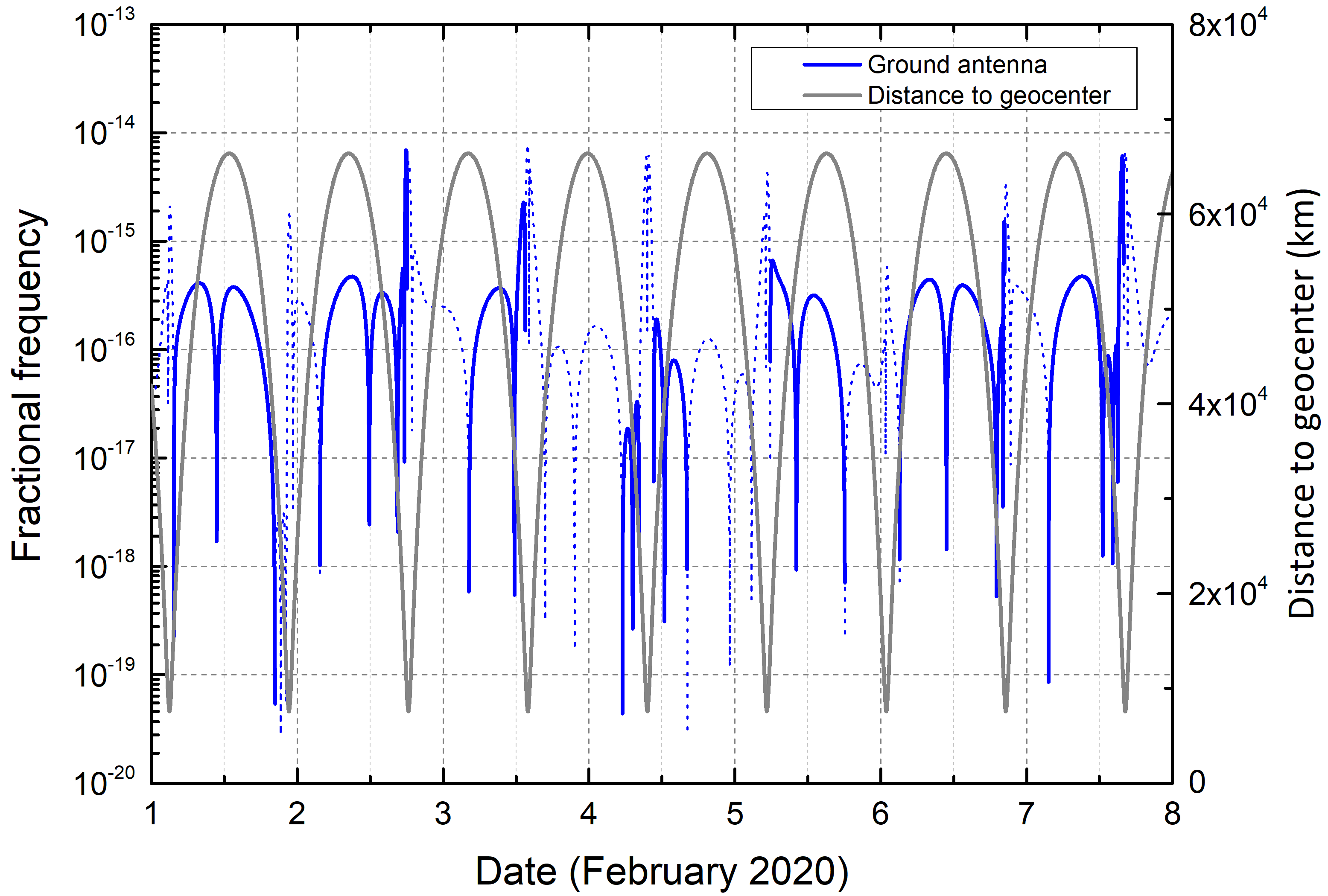}
        \caption{Variation of fractional frequency shift of downlink signal along the orbit due to phase center motion of the Miyun tracking station antenna to the February, 2020. Blue dots indicate the time intervals when the satellite under the horizon for the station.}
        \label{fig:fig7}
\end{figure}

\begin{figure}[h!]              
        \centering
        \includegraphics[scale=0.33]{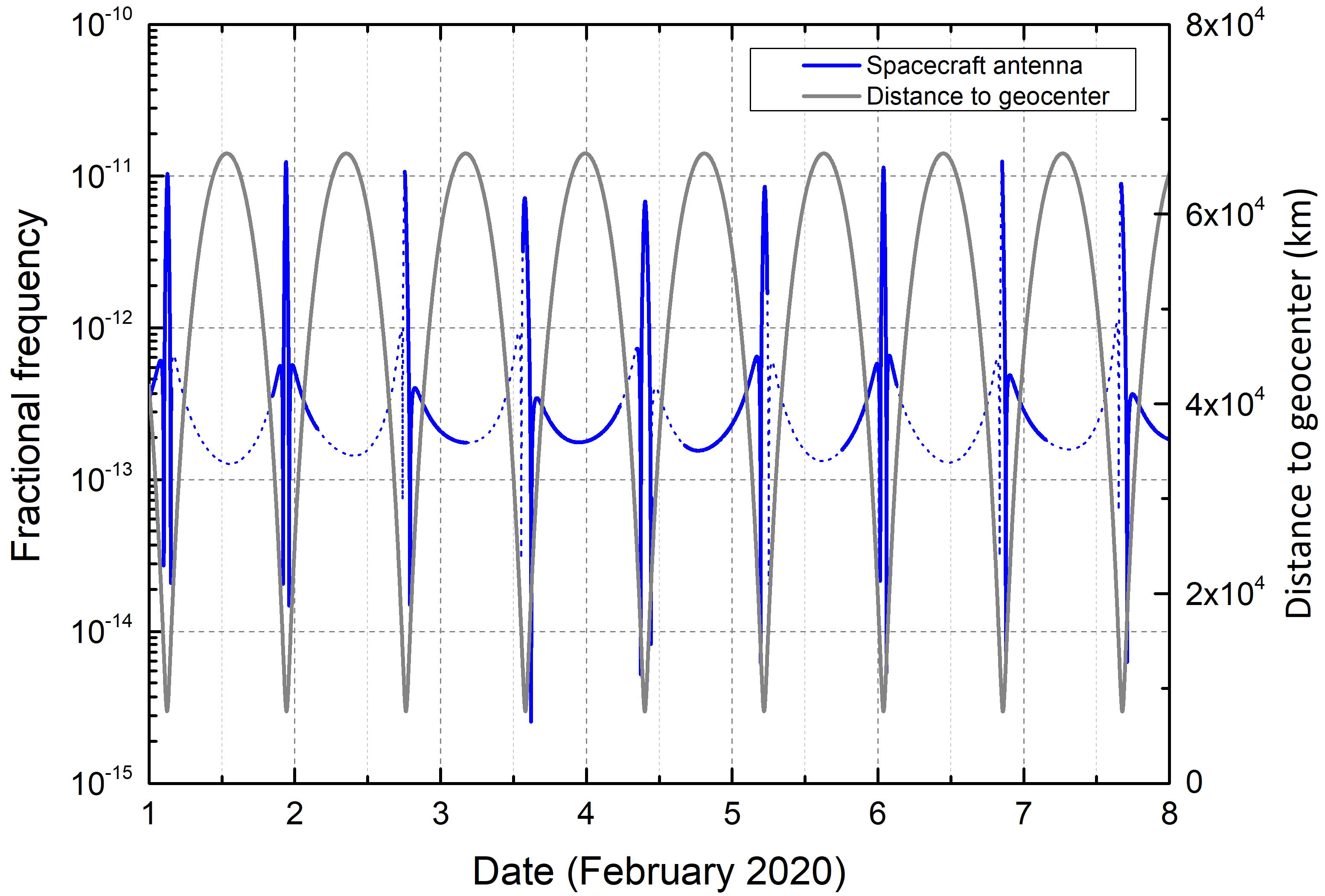}
        \caption{Variation of fractional frequency shift of downlink signal along the orbit due to phase center motion of on-board antenna for the epoch
of February 2020. Blue dots indicate the time intervals when the satellite under the horizon for the station.}
        \label{fig:fig8}
\end{figure}

Now we consider the importance of taking into account the antenna phase center motion effect for orbit determination of spacecrafts of possible future space-VLBI missions, taking as an example the Chinese concept detailed in \cite{hong-2014-aa}. This mission concept implies using two space-based radio telescopes, mounted on satellites with highly elliptic orbits with the perigee and apogee altitudes
of 1,200 and 60,000 km, respectively, inclination of 28.5$^{\circ}$ at first phase of the program. Possible tracking stations include Kaishi, Miyun, Sanya and Shanghai 25 m radio telescope, which can also be functioned as a tracking station after update. For our simulations we take the satellite geometry identical to that of RadioAstron and choose the tracking station of Miyun. Other parameters of the simulation are given in Table~\ref{table:table3}. The Miyun antenna has a very small axis offset of 0.4 cm, thus we expect the effect due to the ground antenna to be negligible, while the effect due to the onboard antenna is expected to be comparable to that of RadioAstron. We see from Figs.~\ref{fig:fig7} and \ref{fig:fig8} that it is indeed the case. The magnitude of the effect for the ground antenna is negligible for orbit determination (less than 0.003 mm/s), while the effect due to the satellite antenna is non-negligible, reaching 3.7 mm/s near perigees.

\begin{table}[h!]               
        \centering
        \begin{tabular}{|l|c|} \hline
                Apogee (km)& 60,000 \\ \hline
                Perigee (km)& 1,200 \\ \hline
                $\theta (^{\circ})$ & 28.5 \\ \hline
                $\varOmega (^{\circ})$ & 15 \\ \hline
                $\omega (^{\circ})$ &  220 \\ \hline
                $M_{0}$ & 0 \\ \hline
                $T_{0}$ & 01/01/2020 00:00:00 UTC\\ \hline
        \end{tabular}
        \caption{Orbital parameters of simulation. Column 1-2: apogee and perigee heights; column 3: inclination angle of the satellite orbital plane; column 4: longitude of the ascending node; column 5: argument of perigee; column 6: mean anomaly at epoch; column 7: epoch. }
        \label{table:table3}
\end{table}

It should be noted that although the ground antenna effect is small for ordinary orbit determination, it cannot be considered small if one intends to perform high-precision experiments, such as measuring the gravitational redshift of an onboard atomic clock to a high accuracy, which is a natural extension of the scientific output of a space-VLBI mission \cite{litvinov-2017-pla}. This case is considered in the next section.

\section{Compensation of the antenna phase center motion effect}

Consider the phase center motion effect for the 2-way mode of the satellite link. For this case the correction should be applied for the up leg of the
link at the time of transmission and for the down leg at the time of reception of the signal:
\begin{equation}        \label{eq:eq3}
        \frac{\Delta f_{\mathrm{pcm}}^{\mathrm{2w}}}{f}
         = \frac{L}{c}\left(\sin{\theta}(t_{3}) \cdot \dot\theta(t_{3}) + \sin{\theta}(t_{1}) \cdot \dot\theta(t_{1})\right)
        -2\:\frac{\mathbf{b} \cdot \dot{\mathbf{s}}(t_{2})}{c}
\end{equation}
where $\Delta t_{21}$ -- signal propagation time in up leg, $\Delta t_{32}$ -- signal propagation time in down leg.
For a simultaneously received signal of the one-way link the correction is:
\begin{align}   \label{eq:eq4}
        \frac{\Delta f_{\mathrm{pcm}}^{\mathrm{1w}}}{f}
        = \frac{L}{c}\sin{\theta}(t_{3}) \cdot \dot\theta(t_{3})
        -\frac{\mathbf{b} \cdot \dot{\mathbf{s}}(t_{2})}{c}
\end{align}

\begin{figure}[h]               
        \centering
        \includegraphics[scale=0.4]{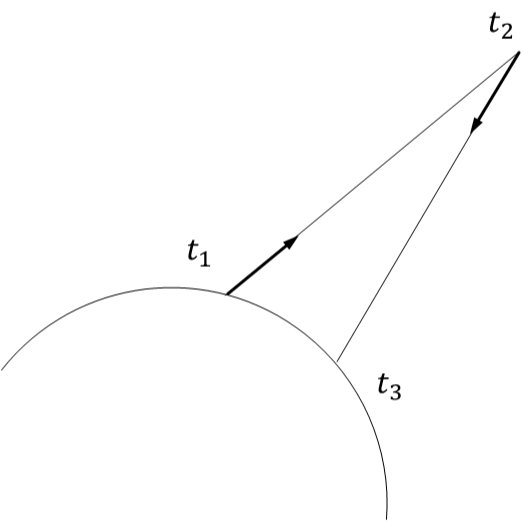}
        \caption{Kinematic characteristics of spacecraft and tracking station.}
        \label{fig:fig10}
\end{figure}

The computed value of the frequency of the one-way signal received from the spacecraft is:
\begin{equation}        \label{eq:eq5}
        \Delta f_{\mathrm{1w}} = \Delta f_{\mathrm{grav}} + \Delta f_{\mathrm{kin}} + \Delta f_{\mathrm{media}} + \Delta f_{\mathrm{pcm}}^{\mathrm{1w}}
\end{equation}          
where $\Delta f_{\mathrm{grav}}$ is the gravitational redshift, $\Delta f_{\mathrm{kin}}$ is kinematic frequency shift, $\Delta f_{\mathrm{media}}$ is media correction, $\Delta f_{\mathrm{pcm}}^{\mathrm{1w}}$ is correction due to the phase center motion in one-way link.
The computed value of the two-way frequency of the signal transmitted by the tracking station, received and coherently retransmitted by the spacecraft, and finally received by the tracking station is:
\begin{equation}        \label{eq:eq6}
        \Delta f_{\mathrm{2w}} = 2\Delta f_{\mathrm{kin}} + 2\Delta f_{\mathrm{media}} + \Delta f_{\mathrm{pcm}}^{\mathrm{2w}}
\end{equation}
where $\Delta f_{\mathrm{pcm}}^{\mathrm{2w}}$ is correction due to the phase center motion in two-way link . Now consider the following observable:
$ \Delta f_{\mathrm{1w}} - \dfrac{\Delta f_{\mathrm{2w}}}{2}  $. 
This observable is important in that its computed value:
\begin{equation}        \label{eq:eq7}
        \Delta f_{\mathrm{1w}} - \frac{\Delta f_{\mathrm{2w}}}{2} 
        = \Delta f_{\mathrm{grav}} 
        + \Delta f_{\mathrm{kin}}^{\mathrm{res}} 
        + \Delta f_{\mathrm{media}}^{\mathrm{res}}
        + \Delta f_{\mathrm{pcm}}^{\mathrm{res}}
\end{equation}
contains the contribution of the gravitational frequency shift of the one-way downlink but cancels, up to non-reciprocity of the up and down legs, the nonrelativistic Doppler effect, which is orders of magnitude larger than all other terms ($\sim10^{-5}$). Thus this observable is very useful in high-precision gravitational experiments, which was first demonstrated by the Gravity Probe A experiment \cite{vessot-levine-1980-prl} and then exploited
in the gravitational redshift experiment with RadioAstron \cite{litvinov-2017-pla}. Alternatively, the left side of Eq.~(\ref{eq:eq7}) can be used for studying the gravitation field at the location of the spacecraft. In the first case one can searches for violations of the Einstein formula for the gravitational frequency shift by comparing the measured value of the observable (\ref{eq:eq6}) with the one computed according to general relativity. In the second application one uses the measured value of the observable to determine the gravitational potential at the location of the spacecraft. 

Note that, although the observables (\ref{eq:eq5}) and (\ref{eq:eq6}) include the contributions of the ground and onboard antenna effects, the  observable (\ref{eq:eq7}) does not:
\begin{equation}        \label{eq:eq8}
        \frac{\Delta f_{\mathrm{pcm}}}{f} 
        = \frac{L}{\:c} \left( \cos\theta(t_{3})\dot\theta^{2}(t_{3}) + \sin\theta(t_{3}) \ddot\theta(t_{3}) \right)\Delta t_{32} 
\end{equation}
The terms due to the onboard antenna is completely cancelled (neglecting the
time delay of the onboard transponder), while the terms due to the ground
antenna are reduced, with the residual term proportional to the signal light travel time, $\Delta t_{32}$, between the spacecraft and the tracking station,
 and depends on the
rate of angular motion, $\dot\theta(t_{3})$, of the spacecraft relative to the tracking station and its rate of change, $\ddot\theta(t_{3})$. 
 Thus, even though mechanically steerable spacecraft dish antennas are naturally not ideal for high-precision measurements, they can be used together with
the compensation scheme of the left side of Eq.~(\ref{eq:eq7}).

\section{Conclusions}

The phase center motion effect for steerable high-gain antennas gives a significant contribution to computed values of observables in orbit determination and VLBI. Although the effect has been known for quite a while, we showed some novel aspects of it. In particular, we demonstrated, based on our experience in orbit determination of the RadioAstron spacecraft, that not only the effect of the ground parabolic antenna phase center motion needs to be taken into account for regular orbit determination but also the effect due to the onboard antenna is almost as large. This conclusion is important for planning of the future space-VLBI missions, an example of which we considered in detail.

We also considered the prospects of performing high-precision space-based experiments with spacecrafts equipped with mechanically steerable high-gain parabolic antennas. We showed that a specific kind of experiments, i.e. those aimed at testing the fundamentals of gravity and also those for chronometric studies of the gravitational field, can safely be performed using  mechanically steerable antennas by exploiting a specific configuration of the satellite
communication links.

\section*{Acknowledgements}

Research for the RadioAstron gravitational redshift experiment is supported by the Russian Science Foundation grant 17-12-01488.
The RadioAstron project is led by the Astro Space Center of the Lebedev Physical Institute of the Russian Academy of Sciences and the Lavochkin Scientific and Production Association under a contract with the Russian Federal Space Agency, in collaboration with partner organizations in Russia and other countries.


\bibliography{antenna_effect_v01}

\end{document}